\documentclass[twocolumn,showpacs,preprintnumbers,amsmath,amssymb]{revtex4}

\usepackage{graphicx}
\usepackage{dcolumn}
\usepackage{bm}

\begin{document}

\preprint{}

\title{In--Flight ($K^-,p$) Reactions for the Formation of Kaonic Atoms and Kaonic Nuclei in Green function method}

\author{J. Yamagata$^1$, H. Nagahiro$^2$}
\author{S. Hirenzaki$^1$}%
\affiliation{$^1$Department of Physics, Nara Women's University, Nara 630-8506, Japan\\
$^2$Research Center for Nuclear Physics (RCNP), Osaka University, Ibaraki, Osaka 567-0047, Japan
}%

\date{\today}

\begin{abstract}
We study theoretically the kaonic atom and kaonic nucleus formations in the in--flight ($K^-,p$) reactions using the Green function method, which is suited to evaluate formation rates both of stable and unstable bound systems. We consider $^{12}$C and $^{16}$O as the targets and calculate the spectra of the ($K^-,p$) reactions. We conclude that no peak structure due to kaonic nucleus formation is expected in the reaction spectra calculated with the chiral unitary kaon--nucleus optical potential. In the spectra with the phenomenological deep kaon--nucleus potential, we may have possibilities to observe some structures due to the formation of the kaonic nucleus states. For all cases, we find clear signals due to the kaonic atom formations in the reaction spectra, which show the very interesting structures like the $'$resonance dip$'$ instead of the $'$resonance peak$'$ for the atomic $1s$ state formation.
\end{abstract}

\maketitle

\section{\label{sec:intro}Introduction}
Kaonic atoms and kaonic nuclei 
carry important information concerning the $K^-$--nucleon interaction in nuclear medium.  This information is very important to know the kaon properties at finite density and, for example, to determine the constraints on kaon condensation in high density matter. In recent years, there have been important developments in the studies of kaonic nuclear states, which are kaon--nucleus bound systems by the strong interaction inside the nucleus. Experimental studies of the kaonic nuclear states using in--flight (${\bar K},N$) reactions were proposed and performed by Kishimoto and his collaborators \cite{Kishimoto99,Kishimoto03}. And the first theoretical results of the energy spectra of the in--flight (${\bar K},N$) reaction were obtained in Ref.~\cite{gata05}. Experiments employing stopped (${\bar K},N$) reactions were carried out by Iwasaki, T. Suzuki and their collaborators and reported in Refs.~\cite{Iwasaki03}. In these experiments, they found some possible indications of the existence of kaonic nuclear states with significantly narrow widths. Another indication of $K^-pp$ bound state was reported by the FINUDA experiment \cite{finuda}. There are also theoretical studies of the structure and formation of kaonic nuclear states related to these experimental activities \cite{Akaishi02}. It should be noted that these theoretical studies predict the possible existence of ultra--high density states in kaonic nuclear systems \cite{Akaishi02,dote04}. A critical analysis of the latest stopped $K$ experimental data was also reported by Oset and Toki~\cite{oset05}. We think that existence of the NARROW kaonic nuclear states is still controversial and we need further studies.

Another very interesting feature of the kaon--nucleus bound systems is based on the fact that the properties 
of kaons in nuclei are strongly influenced by the change undergone by 
$\Lambda(1405)$ in nuclear medium, because $\Lambda(1405)$ is a resonance 
state just below the kaon-nucleon threshold.  In fact, there are studies 
of kaonic atoms carried out by modifying the properties of $\Lambda(1405)$ in nuclear medium \cite{alberg76,wei,miz}. 
These works reproduce the properties of specific kaonic atoms reasonably well. In Ref.~\cite{batty97}, the phenomenological study of kaonic atoms are performed comprehensively, where the density--dependent potentials are considered for $\chi ^2$ fitting to take into account possible non--linear effects which could be due to $\Lambda(1405)$ resonance.

 Recently, there have been significant developments in the description
of hadron properties in terms of the $SU(3)$ chiral Lagrangian. The interpretation of the $\Lambda(1405)$ resonance state as a baryon--meson coupled system proposed in Ref.~\cite{dalitz67} is also supported by the studies with the chiral Lagrangian~\cite{kai,ose}. Subsequently, the properties of $\Lambda(1405)$
in nuclear medium 
using the $SU(3)$ chiral unitary model were also investigated by
Waas et al. \cite{waa}, Lutz \cite{lut}, Ramos and Oset \cite{ram}, and Ciepl$\acute{\rm y}$ et al. \cite{ciep01}. 
All of these works considered the Pauli effect on the 
intermediate nucleons. In addition, in Ref.~\cite{lut}, the self-energy of the 
kaon in the intermediate states is considered, and in Ref.~\cite{ram}, the 
self-energies of the pions and baryons are also taken into account.  These 
approaches lead to a kaon self-energy in nuclear medium that can be 
tested with kaonic atoms and kaonic nuclei. The in--medium ${\bar K}$ properties have been studied also in Refs.~\cite{tolo01,tolo02} based on meson--exchange J$\ddot{\rm u}$lich ${\bar K}N$ interaction, and in Ref.~\cite{schaffner00} for the environment in heavy ion collisions. 

In a previous work \cite{hirenzaki00}, we adopted the scattering amplitude in nuclear 
medium calculated by Ramos and Oset \cite{ram} for studies of
kaonic atoms, and demonstrated the ability to reproduce the kaonic atom data reasonably well. Baca et al.~\cite{baca00} modify the potential by adding the small phenomenological term and improve the fit to the atomic data significantly, which is comparable to the phenomenological fit~\cite{batty97}.
We then calculated the deeply bound kaonic
atoms for $^{16}$O and $^{40}$Ca, which have narrow widths 
and are believed to be observable with well-suited experimental methods. 
We also obtained
very deep kaonic nuclear states, which have large decay widths, 
of the order of several tens of MeV. Similar results were also obtained in Refs. \cite{16.5_Friedman99,Friedman99}.

In this paper, we study the in--flight ($K^-,p$) reactions systematically using the Green function method~\cite{morimatsu85} for the formation of kaon--nucleus bound systems like kaonic atoms and kaonic nuclei. The Green function method is known to be suited for evaluations of formation rates both of stable and unstable bound systems. The ($K^-,p$) reaction for the ${\bar K}$--nucleus system formation was proposed in Refs.~\cite{Kishimoto99} and~\cite{Friedman99}, and first theoretical results of the reaction energy spectra were obtained in Ref.~\cite{gata05} using the same theoretical approach of Ref.~\cite{hirenzaki91} for the deeply bound pionic atom formation reaction~\cite{tok,gilg00}. We use the Green function method in this paper to get more realistic reaction spectra theoretically than our previous results~\cite{gata05} where we used the effective number approach which is best suited for discrete states formation. By the Green function method we can consider the unstable states, which have large widths and have some overlap with neighbor states, as well as stable states, which are isolated well from other states, simultaneously in realistic theoretical formalism~\cite{morimatsu85}. We can also include the contribution from quasi--free kaon production in the final states in this formalism. We think this evaluation is very interesting and important to know the experimental feasibilities of the ($K^-,p$) reaction and to understand the deeper meanings of the observed spectra~\cite{Kishimoto99,Kishimoto03}. We believe that the realistic calculations of the formation spectra are necessary for all observed results to study the kaon properties in nuclear medium, and to get the decisive conclusions.

In Sec.~I\hspace{-.1em}I, we describe the theoretical models of the kaon--nucleus optical potential, and of the ($K^-,p$) reactions. Numerical results are presented and discussed in Sec.~I\hspace{-.1em}I\hspace{-.1em}I for $^{12}$C and $^{16}$O target cases. We give conclusions of this paper in Sec.~I\hspace{-.1em}V.

\section{\label{formalism}Formalism}

We have improved our theoretical formalism in our previous work in Ref.~\cite{gata05} in the following three points;
\begin{enumerate}
\item[(1)] the energy dependence of the kaon--nucleus optical potential is correctly taken into account,
\item[(2)] the Green function method is adopted to calculate the ($K^-,p$) reaction spectra,
\item[(3)] the quasi--free kaon production is evaluated and included in the calculated spectra,
\end{enumerate}
\noindent
as described in detail below.

In order to investigate the in--flight ($K^-,p$) spectra theoretically, we consider the Klein--Gordon equation
\begin{equation}
\label{KG_eq}
[-{\bm \nabla}^2+\mu^2+2\mu V_{\rm opt}(r)]\phi({\bm r})=[\omega-V_{\rm coul}(r)]^2 \phi({\bm r}).
\end{equation}

\noindent
Here, $\mu$ is the kaon--nucleus reduced mass and $V_{\rm coul}(r)$ is the Coulomb potential with a finite nuclear size:
\begin{equation}
\label{V_coul}
V_{\rm coul}(r)=-e^2 \int \frac{\rho_{\rm ch}(r')}{\bigl| {\bm r}-{\bm r'}\bigr|}d^3 r',
\end{equation}
\noindent
where $\rho_{\rm ch}(r)$ is the charge distribution of the nucleus. We employ the empirical Woods--Saxon form for the density as
\begin{equation}
\label{rho_ch}
\rho_{\rm ch}(r)=\frac{\rho_0}{1+\exp[(r-R)/a]},
\end{equation}
\noindent
where we use $R=1.18A^{1/3}-0.48$ fm and $a=0.5$ fm with $A$, the nuclear mass number. To evaluate the kaon--nucleus optical potential, we use the point nucleon density distributions deduced from the $\rho_{\rm ch}$ in Eq.~(\ref{rho_ch}) by the same prescription described in Sect.~4 in Ref.~\cite{nieves93}. The shapes of the density distributions of the proton and neutron are assumed to be same in this paper.

We use the Green function method~\cite{morimatsu85} to calculate the formation cross sections of the ${\bar K}$--nucleus system in the ($K^-,p$) reactions. The details of the application of the Green function method are found in Refs.~\cite{hayano99,klingl99,jido02}. 

The present method starts with a separation of the reaction cross section into the nuclear response function $S(E)$ and the elementary cross section of the $p({\bar K},p){\bar K}$ with the impulse approximation
\begin{equation}
\label{crossS}
\Biglb( \frac{d^2\sigma}{d\Omega dE}\Bigrb)_{A({\bar K},p)(A-1)\otimes {\bar K}}=\Biglb( \frac{d\sigma}{d\Omega}\Bigrb)^{\rm lab}_{p({\bar K},p){\bar K}}\times S(E).
\end{equation}
\noindent
The forward differential cross section of the elementary process $p({\bar K},p){\bar K}$ in the laboratory frame $\displaystyle \bigl(\frac{d\sigma}{d\Omega}\bigr)^{\rm Lab}_{p({\bar K},p){\bar K}}$ is evaluated to be 8.8 mb/sr at $T_K=600~{\rm MeV}$ using the $K^-p$ elastic cross section data in Ref.~\cite{conforto76}. We should mention here that the corrections to this evaluation were reported in Ref.~\cite{ciep01}, which reduce the elementary cross section to be 3.6 mb/sr effectively. In this paper, we show the all calculated results with assuming the elementary cross section to be 8.8 mb/sr.

The calculation of the nuclear response function with the complex potential is formulated by Morimatsu and Yazaki~\cite{morimatsu85} in a generic form as 
\begin{equation}
\label{S(E)}
S(E)=-\frac{1}{\pi} {\rm Im}\sum_f \tau^\dagger_f G(E) \tau_f,
\end{equation}
\noindent
where the summation is taken over all possible final states. The amplitude $\tau_f$ denotes the transition of the incident particle (${\bar K}$) to the proton--hole and the outgoing ejectile ($p$) , involving the proton--hole wavefunction $\psi_{j_p}$ and the distorted waves $\chi_i$ and $\chi_f$, of the projectile and ejectile, taking the appropriate spin sum
\begin{equation}
\label{tau}
\tau_f({\bm r})=\chi_f^*({\bm r})\xi_{1/2,m_s}^*[Y_{l_{\bar K}}^*(\hat{\bm r})\otimes \psi_{j_p}({\bm r})]_{JM}\chi_i({\bm r})~,~
\end{equation}
\noindent
with the meson angular wavefunction $Y_{l_{\bar K}}(\hat{\bm r})$ and the spin wavefunction $\xi_{1/2,m_s}$ of the ejectile. The distorted waves are written with the distortion factor $F({\bm r})$ as 
\begin{equation}
\label{chi}
\chi_f^*({\bm r})\chi_i({\bm r})=\exp(i{\bm q}\cdot{\bm r})F({\bm r})~,~
\end{equation}
\noindent
with the momentum transfer ${\bm q}$. The distortion factor $F({\bm r})$ is defined as
\begin{equation}
\label{distortionfactor}
F({\bm r})=\exp \Bigl( -\frac{1}{2}\bar{\sigma}\int^{\infty}_{-\infty}dz'\bar{\rho}(z',{\bm b})\Bigr)~,~
\end{equation}
where $\bar{\sigma}$ is the averaged distortion cross section defined as,
\begin{equation}
\bar{\sigma}=\frac{\sigma_{{\bar K}N}+\sigma_{pN}}{2}~,
\end{equation}
with the total cross sections of the incident kaon and emitted proton with the nucleons in the nucleus.~The averaged nuclear density $\bar{\rho}(z',{\bm b}$) in Eq.~(\ref{distortionfactor}) is defined as,
\begin{equation}
 \bar{\rho}(r)=\frac{\rho_0}{1+\exp[(r-\bar{R})/\bar{a}]}~,
\end{equation}
in the polar coordinates with the averaged radial parameter $\bar{R}$ and diffuseness parameter $\bar{a}$ defined as,
\begin{equation}
\bar{R}=\frac{R_i+R_f}{2}~,
\end{equation}
and
\begin{equation}
\bar{a}=\frac{a_i+a_f}{2}~,
\end{equation}
with the density parameters of the nuclei in the initial and final states.\\

The Green function $G(E)$ contains the kaon--nucleus optical potential in the Hamiltonian $H_{\bar K}$ as,
\begin{equation}
\label{Green1}
G(E,{\bm r},{\bm r'})=\langle p^{-1}|\phi_{\bar K}({\bm r})\frac{1}{E-H_{\bar K}+i\epsilon}\phi^\dagger_{\bar K}({\bm r})|p^{-1}\rangle~,
\end{equation}
where $\phi^\dagger_{\bar K}$ is the meson creation operator, $|p^{-1}\rangle$ the proton--hole state, and $E$ the kaon energy defined as $E=T_{\bar K}-T_p-S_p$ using the kinematical variables defined in the formation reaction as, $T_{\bar K}$ is the incident kaon kinetic energy, $T_p$ the emitted proton kinetic energy, and $S_p$ the proton separation energy from the each proton single particle level which is compiled in Table~${\rm I\hspace{-.1em}I\hspace{-.1em}I}$ in Ref.~\cite{gata05}.
Obtaining the Green function with the optical potential is essentially the same as solving the associated Klein--Gordon equation.~We can calculate the nuclear response function $S(E)$ from $\tau^\dagger_f({\bm r})G(E;{\bm r},{\bm r'})\tau_f({\bm r'})$ by performing appropriate numerical integrations for the variables ${\bm r}$ and ${\bm r'}$.

In the Green function formalism, we can calculate the response function $S(E)$ for both bound and quasi--free kaon production energy regions, and we can also perform the summation of the kaon final states without assuming the existence of the discrete kaon bound states, which could disappear in the cases with the strongly absorptive optical potential.

As for the kaon--nucleus interaction, we consider two different energy dependent optical potentials, that obtained with the chiral unitary approach \cite{ram}, and that obtained with a phenomenological fit \cite{batty97} with an appropriate phase space factor for the imaginary part \cite{mares04}, which will be explained below. We take into account the energy dependence of both potentials in the present calculation. The optical potentials of the chiral unitary approach, which is obtained by the kaon self--energy in nuclear matter with the local density approximation, is described in detail in Ref.~\cite{ram}, and we use the latest results of the model~\cite{ramp}.

The optical potential obtained in the phenomenological fit~\cite{batty97} is written as,
\begin{equation}
\label{V_opt_ph}
2\mu V_{\rm opt}=-4 \pi \eta a_{\rm eff}(\rho,E)\rho(r),
\end{equation}

\noindent
where $a_{\rm eff}(\rho,E)$ is a density and energy dependent effective scattering length and $\eta=1+m_K/M_N$. The $a_{\rm eff}(\rho,E)$ is parameterized as,
\begin{equation}
\label{real_aeff}
{\rm Re}~a_{\rm eff}=-0.15+1.66(\rho/\rho_0)^{0.24}~{\rm fm},
\end{equation}
and
\begin{equation}
\label{im_aeff}
{\rm Im}~a_{\rm eff}=[0.62-0.04(\rho/\rho_0)^{0.24}] f^{\rm MFG}(E)~{\rm fm}.
\end{equation}

\noindent
The $f^{\rm MFG}(E)$ is a phase space factor introduced to evaluate the phase space volume of decay channels and defined in Ref.~\cite{mares04} by Mare$\check{\rm s}$, Friedman, and Gal as,

\begin{equation}
f^{\rm MFG}(E)=0.8f^{\rm MFG}_1(E)+0.2f^{\rm MFG}_2(E) ,
\label{eq:mfg}
\end{equation}
\noindent
where $f^{\rm MFG}_1$ and $f^{\rm MFG}_2$ are the phase space factors for 
$\bar{K} N \rightarrow \pi \Sigma$ and 
$\bar{K} NN \rightarrow \Sigma  N$ decay, respectively. These factors are defined as 
\begin{widetext}
\begin{equation}
f^{\rm MFG}_1(E)=\frac{M^3_{01}}{M^3_1}\sqrt{\frac{[M^2_1-(m_{\pi}+m_{\Sigma})^2][M^2_1-(m_{\Sigma}-m_{\pi})^2]}
{[M^2_{01}-(m_{\pi}+m_{\Sigma})^2][M^2_{01}-(m_{\Sigma}-m_{\pi})^2]}}
\times \theta(M_1-m_{\pi}-m_{\Sigma}) ,
\label{eq:mfg1}
\end{equation}
\end{widetext}
and
\begin{widetext}
\begin{equation}
f^{\rm MFG}_2(E)=\frac{M^3_{02}}{M^3_2}\sqrt{\frac{[M^2_2-(m_ N+m_{\Sigma})^2][M^2_2-(m_{\Sigma}-m_ N)^2]}
{[M^2_{02}-(m_ N+m_{\Sigma})^2][M^2_{02}
-(m_{\Sigma}-m_ N)^2]}}\theta(M_2-m_{\Sigma}-m_ N) .
\label{eq:mfg2}
\end{equation}
\end{widetext}
Here, the branching ratios of mesic decay and non-mesic decay are 
assumed to be 80$\%$ and 20$\%$ in Eq.~(\ref{eq:mfg}). The masses are defined as 
$M_{01}=m_{\bar K}+m_{ N}$, 
$M_1=M_{01}+E$, $M_{02}=m_{\bar K}+2m_{ N}$, 
$M_2=M_{02}+E$, and $E$ is the kaon energy appeared in Eq.~(\ref{Green1}). In this paper, we introduce this factor to reduce the strength of the absorptive potential of the phenomenological fit, due to the phase space suppression for deeply bound kaonic states. To perform more realistic estimations, we should introduce the energy dependence for each absorptive process separately and we should also consider the medium effects for $\pi$ and $\Sigma$ produced by the kaon absorption. In the present calculations, we have not included these effects for simplicity.

\section{\label{results}Numerical Results}
\begin{figure}[htpd]
\includegraphics[width=6.5cm,height=8cm]{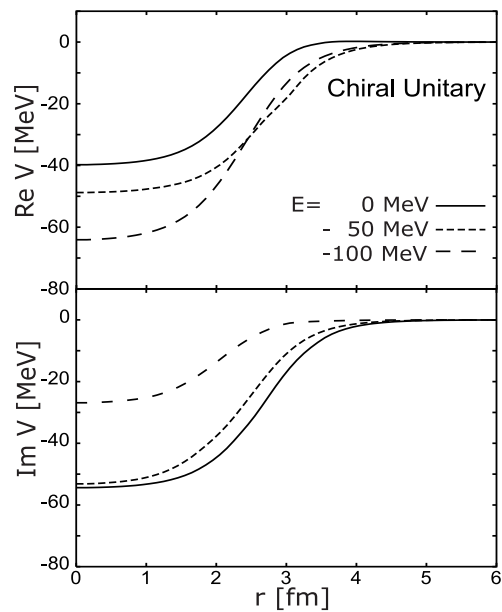}
\caption{\label{fig:pot_chi}The kaon--nucleus optical potential of the chiral unitary approach \cite{ram,ramp} for $^{11}$B as a function of the radial coordinate $r$. The upper and lower panels show the real and imaginary parts, respectively. The solid, dotted and dashed lines indicate the potential strength for the kaon energies $E=0~{\rm MeV}, E=-50~{\rm MeV},~{\rm and}~E=-100~{\rm MeV}$, respectively.}
\end{figure}

\begin{figure}[htpd]
\includegraphics[width=6.5cm,height=8cm]{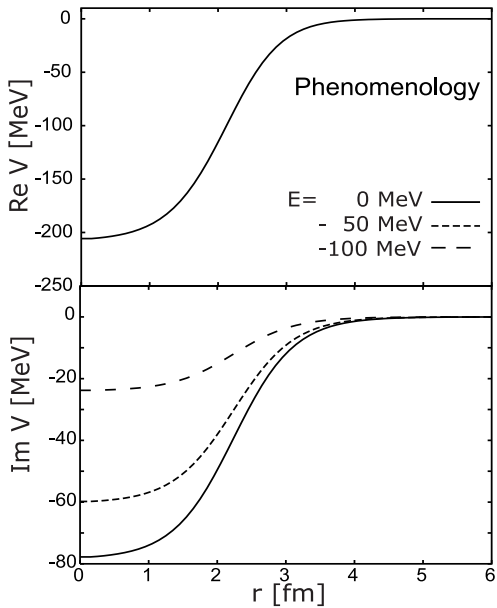}
\caption{\label{fig:pot_ph}The kaon--nucleus optical potential of the phenomenological fit \cite{batty97} with the phase space factor \cite{mares04} for $^{11}$B as a function of the radial coordinate $r$. The upper and lower panels show the real and imaginary parts, respectively. The solid, dotted and dashed lines indicate the potential strength for the kaon energies $E=0~{\rm MeV}, E=-50~{\rm MeV},~{\rm and}~E=-100~{\rm MeV}$, respectively.~The real part does not have the energy dependence.}
\end{figure}

In the present paper, we take into account the energy dependence of the optical potentials correctly. We show the kaon--nucleus optical potentials for the $^{11}$B case in Fig.~\ref{fig:pot_chi} for the chiral unitary model and in Fig.~\ref{fig:pot_ph} for the phenomenological fit. Because the real part of the optical potential changes its sign at a certain nuclear density at $E=0$ MeV in both the chiral unitary and phenomenological models, the kaon--nucleus optical potential is attractive, while keeping the repulsive sign for the kaon--nucleon scattering length in free space at $E=0$ MeV.

The real part of the optical potential of the chiral unitary model                                             has the energy dependence as shown in the upper panel in Fig.~\ref{fig:pot_chi} and the potential depth varied from around $-40$ MeV to $-65$ MeV for the kaon energies $0$ MeV~--~$-100$ MeV. As for the imaginary part, the strength of the nuclear absorption reduced from around $-55$ MeV to $-28$ MeV for the kaon energies $0$ MeV~--~$-100$ MeV mainly due to the threshold effects, which is naturally included in the coupled channel calculation in the chiral unitary model~\cite{ram,ramp}.

\begin{figure}[htpd] 
\includegraphics[width=8cm,height=5cm]{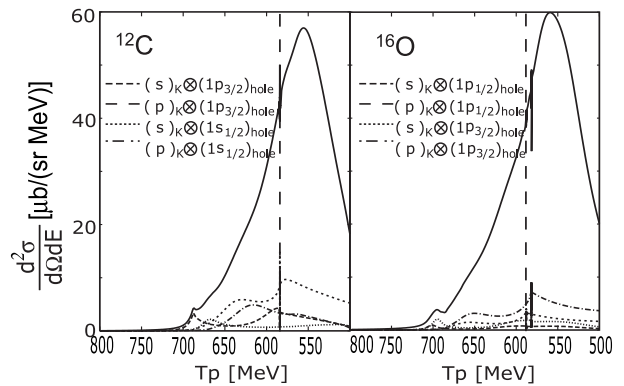}
\caption{\label{fig:omega}Calculated spectra of the $^{12}$C($K^-,p$) and $^{16}$O($K^-,p$) reactions at $T_{K^-}=600$ MeV plotted as functions of the emitted proton energy $T_p$ at $\theta^{\rm lab}_p=0$ (deg.) for the phenomenological optical potential with different energy dependence defined in Eq.~(\ref{eq:omega}) (see text in detail). Each subcomponent is indicated in the figure. The vertical dashed lines indicate the kaon production threshold.}
\end{figure}

As shown in Fig.~\ref{fig:pot_ph}, the real part of the phenomenological optical potential does not have the energy dependence since the parameters are determined by the experimental data of kaonic atoms which have the information only around $E\sim0$ MeV in the present energy scale.~The effects of the possible energy dependence of the real part will be discussed later in Fig.~\ref{fig:omega}. The depth of the real potential of the phenomenological fit is much deeper than that of the chiral unitary model. There are also phenomenological potentials which provide the similar depth with the chiral unitary potential. However, we consider deeper phenomenological potential here because we can expect to have larger possibilities to form deep kaonic nuclear states, and expect to have deeper insight for the kaon bound states by comparing the results obtained from much different potentials. The imaginary part of the optical potential of the phenomenological fit has the energy dependence due to the threshold effects of the decay channels, which are evaluated in Eqs.~(\ref{eq:mfg})--(\ref{eq:mfg2}). The imaginary part of the both potentials have the qualitatively similar depths and similar energy dependence.

In order to calculate the formation spectra of the kaon--nucleus bound systems in the ($K^-,p$) reactions, we use the Green function method~\cite{morimatsu85} in this paper as described in Sec.~\ref{formalism}. The Green function method is suited to evaluate the formation rates both of the stable and unstable systems, and is suited to include simultaneously the contribution of the quasi--free kaon production in the final states to the spectra. For these two reasons, the Green function method is superior to the effective number approach which we have used in the previous paper~\cite{gata05}. On the other hand, the effective number approach is better suited to calculate the formation spectra of the stable states with narrow widths than the Green function method because of much better efficiency for getting the numerical results.

\begin{figure}[htpd]
\includegraphics[width=6.5cm,height=8cm]{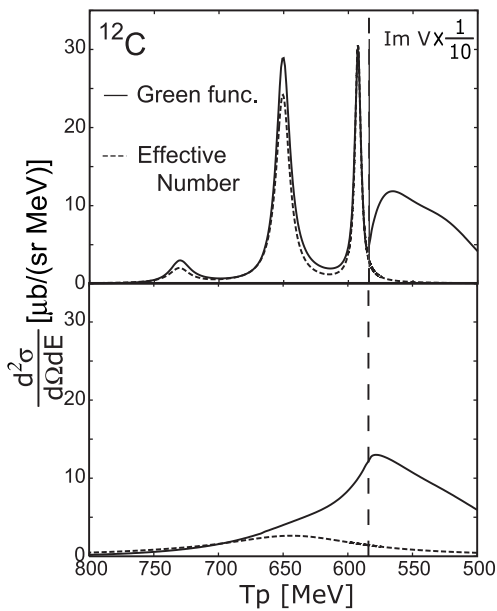}
\caption{\label{fig:GvsE}Subcomponents of the kaonic nucleus formation cross sections in the $^{12}$C($K^-,p$) reactions at $T_{K^-}=600~{\rm MeV}$ are plotted as functions of the emitted proton energy $T_p$ at $\theta^{\rm lab}_p=0$ (deg.) for the phenomenological optical potential Eq.(\protect\ref{V_opt_ph}).~The upper panel shows the results with the reduced imaginary potential $\displaystyle{\frac{1}{10}{\rm Im}V_{\rm opt}}$, while the lower panel shows those with Eq.~(\ref{V_opt_ph}).~The solid lines show the results obtained by the Green function method for the sum of the subcomponents $[(s)_K\otimes (1p_{3/2})^{-1}_p]$ and $[(p)_K\otimes [(1p_{3/2})^{-1}_p]$. The dashed lines show those by the effective number approach for the sum of the subcomponents $[(1s_{\rm nucl})_K\otimes (1p_{3/2})^{-1}_p]$, $[(2p_{\rm nucl})_K\otimes (1p_{3/2})^{-1}_p]$ and $[(2s_{\rm nucl})_K\otimes (1p_{3/2})^{-1}_p]$.~The vertical dashed line indicates the kaon production threshold.}
\end{figure}

In Fig.~\ref{fig:GvsE}, we show the calculated spectra by the Green function method and by the effective number approach for the exactly same cases to show clearly the differences of the both methods. The incident kaon energy is fixed to $T_K=600~{\rm MeV}$, which corresponds to the in--flight experiment~\cite{Kishimoto99,Kishimoto03}.~In the upper panel, we show the calculated results with the phenomenological optical potential Eq.~(\ref{V_opt_ph}) with the reduced strength of the imaginary part $\displaystyle{\frac{1}{10}{\rm Im}V_{\rm opt}}$. We find clearly that each peak corresponding to a bound state has only small width and is well isolated.
In this case, both methods provide quite similar formation spectra for the bound states formation and are considered to be equivalent to each other.~For the quasi--free kaon production energy region, we can see that the quasi--free contributions to the ($K^-,p$) spectra are automatically included in the Green function method as described in Sec.~\ref{formalism}.~The quasi--free contributions are not included in the results of the effective number approach.

In the lower panel, we show the calculated results by both methods for the phenomenological optical potential Eq.~(\ref{V_opt_ph}). We find that there is no peak structure in the spectra corresponding to the bound states formation because of the large widths of the states.
In this case, the spectra calculated by both methods provide significantly different shapes even in the energy region of the kaon bound states. Hence, the Green function method and the effective number approach are not equivalent to each other for the formation of the states with large widths. This discrepancy between the theoretical models is a natural consequence of the fact that the existence of well--isolated eigenstates is assumed in the effective number approach formalism. We use the Green function method in this paper to get the ($K^-,p$) spectra for the formation of kaon bound states which are expected to have large widths.

\begin{figure}[htpd]
\includegraphics[width=8.0cm,height=8cm]{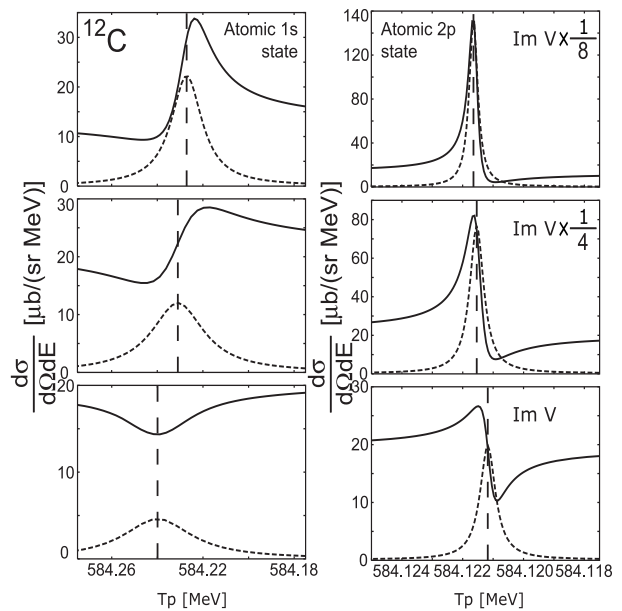}
\caption{\label{fig:atomicstate}Subcomponents of the kaonic atom formation cross sections in the $^{12}$C($K^-,p$) reactions at $T_{K^-}=600$ MeV are plotted as functions of the emitted proton energy $T_p$ at $\theta^{\rm lab}_p=0$ (deg.) for the chiral unitary optical potential~\cite{ram,ramp}. Left and right panels show the calculated results at energies of kaonic atom $1s$ and $2p$ states formation, respectively. The upper and middle panels show the results with the reduced imaginary potential $\displaystyle{\frac{1}{8}}$Im$V_{\rm opt}$ and $\displaystyle{\frac{1}{4}}$Im$V_{\rm opt}$, while the lower panels show those with the full chiral unitary potential. The solid lines show the results obtained by the Green function method for the sum of the subcomponents $[(s)_K\otimes (1p_{3/2})^{-1}_p]$ and $[(p)_K\otimes (1p_{3/2})^{-1}_p]$. The dashed lines show those by the effective number approach for the sum of subcomponents $[(1s_{\rm atom})_K\otimes (1p_{3/2})^{-1}_p]$ and $[(2p_{\rm atom})_K\otimes (1p_{3/2})^{-1}_p]$. The vertical dashed lines indicate the binding energies of kaonic atom $1s$ state (left panels) and $2p$ state (right panels). }
\end{figure}

We also show the comparison of the numerical results with the Green function method and the effective number approach for atomic $1s$ and $2s$ states formation in Fig.~\ref{fig:atomicstate}. In the upper panels, we show the results with reduced strength of the imaginary potential $\displaystyle{\frac{1}{8}}{\rm Im}V_{\rm opt}$, where we can see that the spectra obtained with the both methods are qualitatively same for both atomic states. However, for larger imaginary potentials, the results with both methods show the significant differences as shown in middle and lower panels in Fig.~\ref{fig:atomicstate}. Actually we can find very interesting, even peculiar structures in the results with the full chiral unitary optical potential. Especially, the atomic $1s$ state formation makes the $'$resonance dip$'$ instead of the $'$resonance peak$'$ in the spectrum. Similar results are expected for the phenomenological potential. By all these results show in Fig.~\ref{fig:atomicstate}, we think the origin of the interesting structures for atomic states formation is the interference with the kaon absorptive processes induced by the large imaginary potential.

\begin{figure}[htpd]
\includegraphics[width=6.5cm,height=7cm]{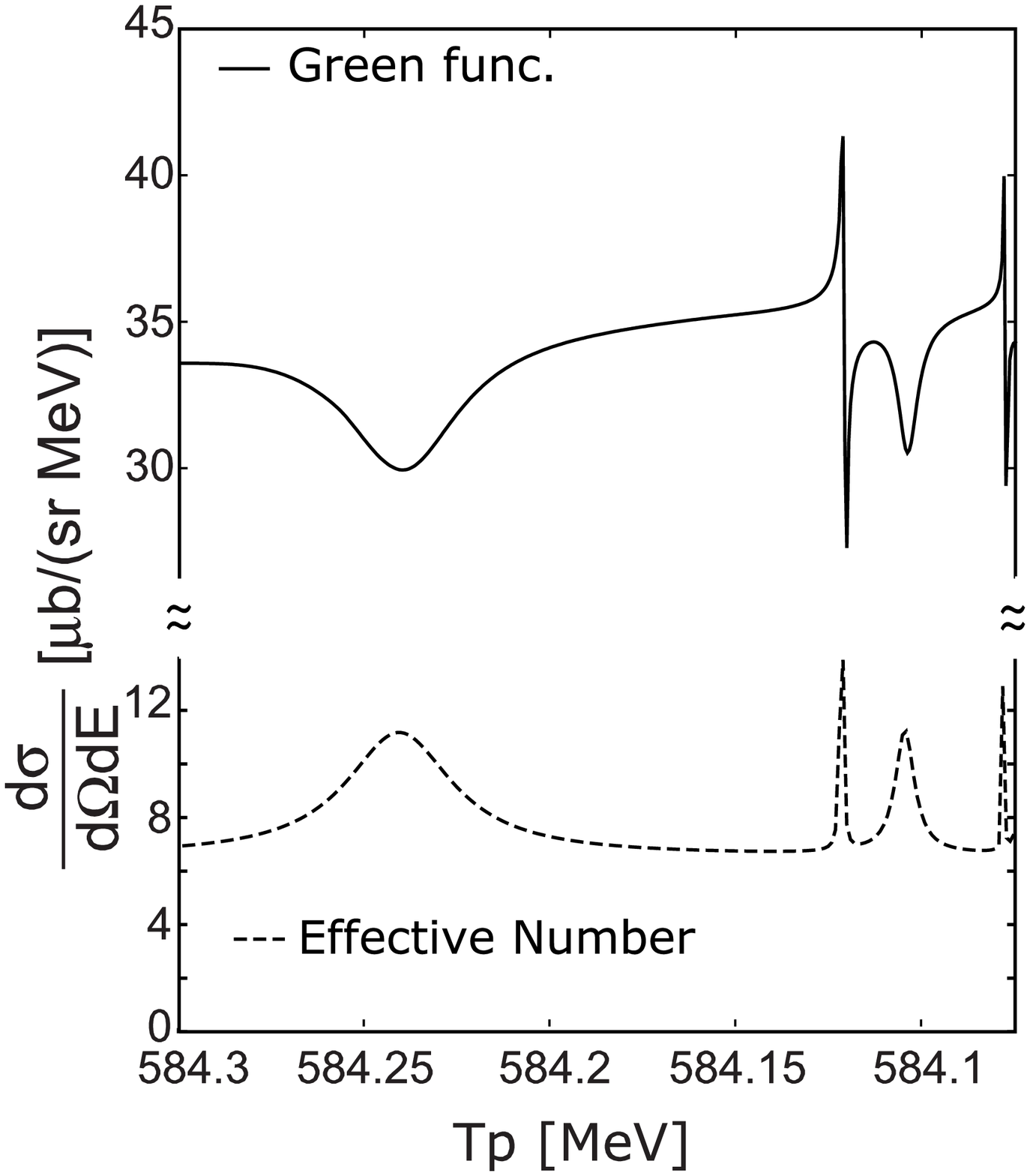}
\caption{\label{fig:atom_Eoff}Kaonic atom formation cross sections in $^{12}$C($K^-,p$) reactions at $T_{K^-}=600$ MeV plotted as functions of the emitted proton energy $T_p$ at $\theta_p^{\rm lab}=0$ (deg.) for the chiral unitary optical potential~\cite{ram,ramp}. Solid line indicates the result of the Green function method and dashed line that of the effective number approach.}
\end{figure}

In Fig.~\ref{fig:atom_Eoff}, we show the fine structure of the total $^{12}$C($K^-,p$) spectra around the kaon production threshold, where we can observe the contributions of atomic states formation.~Here, we have used the optical potential of the chiral unitary model, and we confirm numerically that the spectrum shape is almost the same as that calculated with the phenomenological optical potential in this energy region.~We find again that the spectrum around atomic states formation region is very interesting, where the resonance contributions do not make simple peak structures but make complicated structures as shown by the solid line in Fig.~\ref{fig:atom_Eoff} because of the large imaginary potential.~For comparison, we also show the calculated results by the effective number approach by dashed line, which show the simple peak structures.
We think that it is very interesting to observe experimentally the spectrum shape for kaonic atom formation region.~We mention that these deeply bound kaonic atom states have not been observed yet.

\begin{figure}[htpd]
\includegraphics[width=6.5cm,height=5cm]{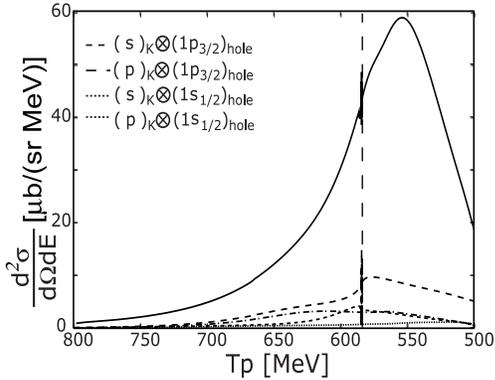}
\caption{\label{fig:Green_ph_Eoff}Calculated spectra of the $^{12}$C($K^-,p$) reaction at $T_{K^-}=600~{\rm MeV}$ 
plotted as functions of the emitted proton energy $T_p$ at $\theta_p^{\rm lab}=0$ (deg.). The energy independent phenomenological potential evaluated at $E=0$ in Eq.~(\ref{V_opt_ph}) is used. Each subcomponent is indicated in the figure.~The vertical dashed line indicates the kaon production threshold.}
\end{figure}

We show firstly the calculated $^{12}$C($K^-,p$) spectrum in Fig.~\ref{fig:Green_ph_Eoff} for the energy independent phenomenological optical potential evaluated at $E=0$ in Eq.~(\ref{V_opt_ph}).~We also show the subcomponents of the spectrum in the same figure.~Though there exist a few kaonic nuclear states in this case, we find that we can not see any peak structures in the ($K^-,p$) spectra because of the large widths.
We also find that the subcomponents of the substitutional states do not dominate the total spectrum because of the finite momentum transfer of this reaction.

\begin{figure}[htpd] 
\includegraphics[width=8cm,height=8cm]{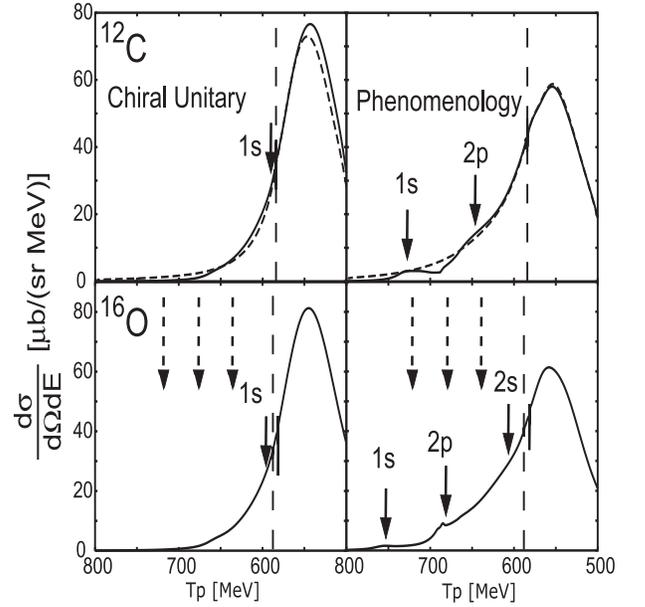}
\caption{\label{fig:Eon_Eoff}Calculated spectra of the $^{12}$C($K^-,p$) (upper panels) and the $^{16}$O($K^-,p$) (lower panels) reactions at $T_{K^-}=600$ MeV plotted as functions of the emitted proton energy $T_p$ at $\theta_p^{\rm lab}=0$ (deg.) for the chiral unitary optical potential (left panels) and the phenomenological optical potential (right panels). Solid lines show the results with the energy dependent potentials and dashed lines those with the energy independent potentials fixed at $E=0$ MeV. The solid arrows indicate the calculated binding energies of the kaonic nuclear states coupled to the ground states of the final state nucleus, namely $(1p_{3/2})^{-1}_p$ state for C target case and $(1p_{1/2})^{-1}_p$ state for O target case, respectively. The binding energies are calculated with the energy dependent potential. The dashed arrows in lower panels indicate the binding energies of the kaonic nuclear states suggested by the experimental group~\cite{Kishimoto99,Kishimoto03}.~The vertical dashed line indicates the kaon production threshold.}
\end{figure}

In order to see the effects of the energy dependence of the potential on the ($K^-,p$) spectra, we show the calculated spectra with energy dependent optical potential in Fig.~\ref{fig:Eon_Eoff} (upper panels) and compare those with energy independent optical potential fixed at $E=0$ for $^{12}$C target case.~The energy dependence of the optical potential is expected to be important since one can think that the strength of the imaginary part of the optical potential could be reduced significantly because of the phase space suppressions of the decay channels, and that the widths of the kaonic nuclear states could be so narrow that one can see the peak structures in the ($K^-,p$) spectra.~As can be seen in the figure, the effect of the energy dependence of the optical potential is tiny for the chiral unitary model and the shape of the ($K^-,p$) spectrum is almost unchanged.
As for the phenomenological optical potential, we can find there appear some small structures around $T_p=650\sim 750~{\rm MeV}$ in the spectrum.~The calculated binding energies corresponding to the kaonic nuclear states are also indicated by solid arrows in the figure for the energy dependent potentials.~This structure could be observed in experiments, though it seems rather difficult.
We should mention here that the small structures in the spectrum show the indications of the existence of the kaonic nuclear states. However, we should be careful to deduce the properties of the kaonic nuclear states from the observed spectra because the shapes and the positions of the structure could be changed from those corresponding to the binding energies and widths of the kaonic states by the energy dependence of the optical potential and by the contributions from other subcomponents. Thus, we think that we need the help of the realistic theoretical calculations of the spectrum to deduce the properties of the kaonic nuclear states from the observed spectrum correctly.

In Fig.~\ref{fig:Eon_Eoff} (lower panels), we show the results for $^{16}$O target case.~The experimental data for this case were reported in Ref.~\cite{Kishimoto03}.
We show calculated results both with the chiral unitary potential and the phenomenological potential.~The dashed arrows in the figure indicate the energies of the possible peak structures suggested by the experimental group~\cite{Kishimoto03}.~We also showed the calculated binding energies by the solid arrows.~We find that our calculated results do not have peak structures at the energies suggested by experiment~\cite{Kishimoto03}.~At present exploratory level, it is rather difficult to deduce conclusions from this comparison because $({\rm i})$ the accuracy of the experimental data seems not enough, and $({\rm i}\hspace{-.1em}{\rm i})$ the effects of nuclear structure changes are not included in the present calculations, which could be important even for C and O as discussed in Ref.~\cite{mares04}.~Hence, in order to perform meaningful comparison between the data and the calculated results, we need accurate data such that we could distinguish the left (chiral unitary) and right (phenomenology) figures in Fig.~\ref{fig:Eon_Eoff} and that we could observe the existence (or non--existence) of the small structures appearing in the spectra obtained with the phenomenological potential. And we also need better theoretical results which incorporate the nuclear structure change effects in order to deduce decisive information on kaonic nuclear states.

\begin{figure}[htpd]
\includegraphics[width=6.5cm,height=5cm]{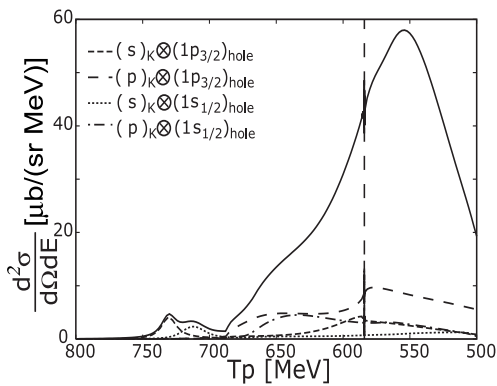}
\caption{\label{fig:Gal0901}Calculated spectra of the $^{12}$C($K^-,p$) reaction at $T_{K^-}=600$ MeV plotted as functions of the emitted proton energy $T_p$ at $\theta_p^{\rm lab}=0$ (deg.) for the energy dependent phenomenological optical potential with the different branching ratio of the decay channels defined in Eq.~(\ref{eq:f_0901}) (see text in detail). Each subcomponent is indicated in the figure.~The vertical dashed line indicates the kaon production threshold.}
\end{figure}

Finally, to see the dependence of the spectrum shape on the assumptions in the theoretical calculations with the phenomenological optical potential, we show in Fig.~\ref{fig:Gal0901} the calculated results using the phase space factor with different branching ratios of the decay channels as,
\begin{equation}
\label{eq:f_0901}
f^{\rm MFG}(E)=0.9f^{\rm MFG}_1(E)+0.1f^{\rm MFG}_2(E)~.~
\end{equation}
Here, we assume the branching ratios of mesic decay and non--mesic decay to be 90$\%$ and 10$\%$.~We find that the structures in the spectrum due to the kaonic nuclear states can be seen clearer around $T_p=700\sim750~{\rm MeV}$ than those in the right panels in Fig.~\ref{fig:Eon_Eoff}.

Furthermore, we show the calculated results with the phenomenological optical potential by replacing the potential term in the Klein--Gordon equation Eq.~(\ref{KG_eq}) as 
\begin{equation}
2\mu V_{\rm opt} \rightarrow 2\omega V_{\rm opt}~,
\label{eq:omega}
\end{equation}
to consider the another type of the energy dependence. Due to the additional energy dependence, the effects of the optical potential are smaller for both real and imaginary parts in the deeper bound region because of the smaller value of the factor $'2\omega'$ in Eq.~(\ref{eq:omega}). Namely, the effective potential strength is small for deeply bound region, which means that the potential is less attractive and less absorptive in smaller $\omega$ region effectively. Hence, as we can see in Fig.~\ref{fig:omega}, the strength of the calculated ($K^-,p$) spectra is smaller for deeper bound region than the results in Fig.~\ref{fig:Eon_Eoff} (right panels). We can find small structures again in the spectra as the indications of the kaonic nuclear states.

\section{\label{conclusions}Conclusions}
We have calculated the in--flight ($K^-,p$) reaction spectra systematically for the formation of kaon--nucleus bound systems by the Green function method.~We have used two different kaon--nucleus optical potentials and compared the calculated results.~One is the optical potential of the chiral unitary model which is based on a microscopic theory and is proved to reproduce the existing atomic data reasonably well.~The energy dependence of the chiral unitary potential is automatically included in the theoretical framework.~The other one is the phenomenological optical potential which is obtained by the $\chi^2$ fitting to the kaonic atom data and has a much deeper real part than that of the chiral unitary potential.~The energy dependence of the phenomenological potential is also considerd.~We should mention here that both potentials provide the qualitatively same energy spectra for the kaonic atoms though they have much different potential depths in the real part.~Thus, it is extremely interesting to compare the calculated results for kaonic nuclear states with these potentials.

The calculated results newly reported in this paper are superior to our previous results reported in Ref.~\cite{gata05} in 3 points as described in Sec.~${\rm I\hspace{-.1em}I}$. Namely, (1) energy dependence of the potential, (2) Green function method, and (3) quasi--free kaon production contributions are considered. We have shown numerically that the effective number approach used in Ref.~\cite{gata05} provides the equivalent results for bound states formation spectra as those provided by the Green function method~\cite{morimatsu85} in cases with the less absorptive optical potentials.~If the potential is so strongly absorptive that resonance peaks overlap with each other, we need to use the Green function method~\cite{morimatsu85}.

We have calculated the ($K^-,p$) spectra systematically on $^{12}$C and $^{16}$O targets at $T_{K^-}=600~{\rm MeV}$.~Our results are summarized as follows; 
\begin{enumerate}
\item[(1)]the ($K^-,p$) spectra calculated with the energy independent kaon--nucleus optical potentials do not show any peak structures indicating the kaonic nuclear states formation, even if such states exist, because of the large widths of the states (Fig.~\ref{fig:Green_ph_Eoff}),
\item[(2)]the ($K^-,p$) spectra calculated with the chiral unitary optical potential do not show any structures indicating the kaonic nuclear states formation in all cases calculated in this paper (Fig.~\ref{fig:Eon_Eoff} left panels),
\item[(3)]the ($K^-,p$) spectra calculated with the energy dependent optical potentials of the phenomenological model Eq.~(\ref{V_opt_ph}) show the small structures around the energy region of the kaonic nuclear states (Fig.~\ref{fig:Eon_Eoff} right panels),
\item[(4)]the small structures mentioned in (3) are not necessarily corresponding to the correct binding energies of the kaonic nuclear states.~The energy dependence of the imaginary part of the optical potential and the contributions of many subcomponents can shift the observed positions of the small structures from those corresponding to the binding energies of the kaonic states, 
\item[(5)]the calculated ($K^-,p$) spectra do not show peak structures at energies suggested by the experimental group \cite{Kishimoto03} for either kaon--nucleus optical potential cases (Fig.~\ref{fig:Eon_Eoff} lower panels).~We think that we need more accurate data and better theoretical results incorporating the effects of nuclear structure changes due to the existence of a kaon in order to make decisive conclusions from this observation,
\item[(6)]the calculated ($K^-,p$) spectra have certain dependence on input parameters like branching ratios of decay channels (Fig.~\ref{fig:Gal0901}), and on the assumed functional form of the energy dependence of the potential (Fig.~\ref{fig:omega}) in the phenomenological potential case,
\item[(7)]the spectra for the kaonic atom states formation show very interesting behavior like the $'$resonance dip$'$ for the $1s$ state formation because of the interference with the absorption processes (Fig.~\ref{fig:atomicstate} and \ref{fig:atom_Eoff}),
\item[(8)]the elementary cross section in Eq.~(\ref{crossS}) is assumed to be 8.8 mb/sr in all calculated results shown in this paper. The corrections to this value are reported in Ref.~\cite{ciep01}, which reduce the elementary cross section to be 3.6 mb/sr. When we take this value, all the calculated cross sections in this paper have to be reduced by roughly a factor of 2.
\end{enumerate}

By considering all results described above, we conclude that we need first to have new experimental data which are so accurate that we can identify the existence (or non--existence) of the small structures appearing in the theoretical calculations in the kaonic nuclear states energy region.~Then, if it exists, we must investigate the origins of the structures very carefully in order to conclude the existence (or non--existence) of the kaonic nuclear states decisively in the studies of the in--flight ($K^-,p$) reactions on the targets like C and O.

Finally, we would like to add a few words for the perspectives in formation reactions of kaonic nuclear states. The stopped ($K^-,p$) reactions on the targets like C and O will provide the similar information as the in--flight reaction cases on the same targets which are reported in this paper. As shown in Refs.~\cite{Iwasaki03, finuda,Akaishi02,dote04}, it may be interesting to consider lighter nuclei as targets. In any cases, we need to evaluate the effects of the nuclear structure changes to the reaction cross sections, which have not been included in our calculations so far. The possibilities of the significant changes of the nuclear structure due to kaon were reported not only for lighter nuclei~\cite{Akaishi02,dote04} but also for heavier nuclei like C and O~\cite{mares04}. The possible changes of the nuclear structures may be expected to reduce the formation cross sections of kaonic nuclear states significantly because of the suppression due to the nuclear transition form factor. On the other hand, a kaon wavefunction confined in smaller spatial dimension is expected to be able to bear the larger momentum transfer in the ($K^-,p$) reactions, and helps to make the cross sections larger. Thus, the quantitative evaluation is quite necessary in order to conclude the size of the observed formation cross sections in Refs.~\cite{Iwasaki03} and the predicted changes of the nuclear structure in Refs.~\cite{Akaishi02,dote04} are really consistent with each other or not. 

We hope that our theoretical results stimulate the experimentalists, and new reliable experimental information will be obtained in the near future in new facilities like J--PARC.
\begin{acknowledgments}
We acknowledge E. Oset and A. Ramos for providing us their calculated results of kaon self--energy at finite densities in the chiral unitary model and for stimulating discussions on kaon bound systems.~We also thank E. Oset and A. Ramos for their careful reading of our manuscript and useful comments. We would like to thank H. Toki, A. Hosaka, Y. Akaishi, Y. Abe, A. Ohnishi and T. Kishimoto for valuable discussions.
This work is partly supported by Grants--in--Aid for scientific research of MonbuKagakusho and Japan Society for the Promotion of Science (No. 16540254), and by Nara Women's University Intramural Grant for Project Research.
\end{acknowledgments}

\end{document}